\documentclass{osa-article}

\journal{oe}

\articletype{Research Article}

\usepackage{graphicx}
\usepackage{dcolumn}
\usepackage{bm}

\usepackage[utf8]{inputenc}
\usepackage[T1]{fontenc}
\usepackage{etoolbox}
\usepackage{multirow}
\usepackage{pifont}
\usepackage[symbol]{footmisc}

\usepackage{amsmath}

\usepackage{amssymb}
\usepackage{booktabs}

\usepackage{lineno}

\begin{document}

\title{Key frames assisted hybrid encoding for photorealistic compressive video sensing}

\author{Honghao Huang,\authormark{1,2} Jiajie Teng,\authormark{1,2} Yu Liang,\authormark{1,2} Chengyang Hu,\authormark{1,2} Minghua Chen,\authormark{1,2} Sigang Yang,\authormark{1,2} and Hongwei Chen,\authormark{1,2,*}}

\address{\authormark{1}Department of Electronic Engineering, Tsinghua University, Beijing, China\\
\authormark{2}Beijing National Research Center for Information Science and Technology (BNRist), Beijing, China}

\email{\authormark{*}chenhw@tsinghua.edu.cn} 



\begin{abstract}
Snapshot compressive imaging (SCI) encodes high-speed scene video into a snapshot measurement and then computationally makes reconstructions, allowing for efficient high-dimensional data acquisition. Numerous algorithms, ranging from regularization-based optimization and deep learning, are being investigated to improve reconstruction quality, but they are still limited by the ill-posed and information-deficient nature of the standard SCI paradigm. To overcome these drawbacks, we propose a new key frames assisted hybrid encoding paradigm for compressive video sensing, termed KH-CVS, that alternatively captures short-exposure key frames without coding and long-exposure encoded compressive frames to jointly reconstruct photorealistic video. With the use of optical flow and spatial warping, a deep convolutional neural network framework is constructed to integrate the benefits of these two types of frames. Extensive experiments on both simulations and real data from the prototype we developed verify the superiority of the proposed method.
\end{abstract}

\section{Introduction}
\begin{figure}[htb]
\begin{center}
\includegraphics[width=0.7\linewidth]{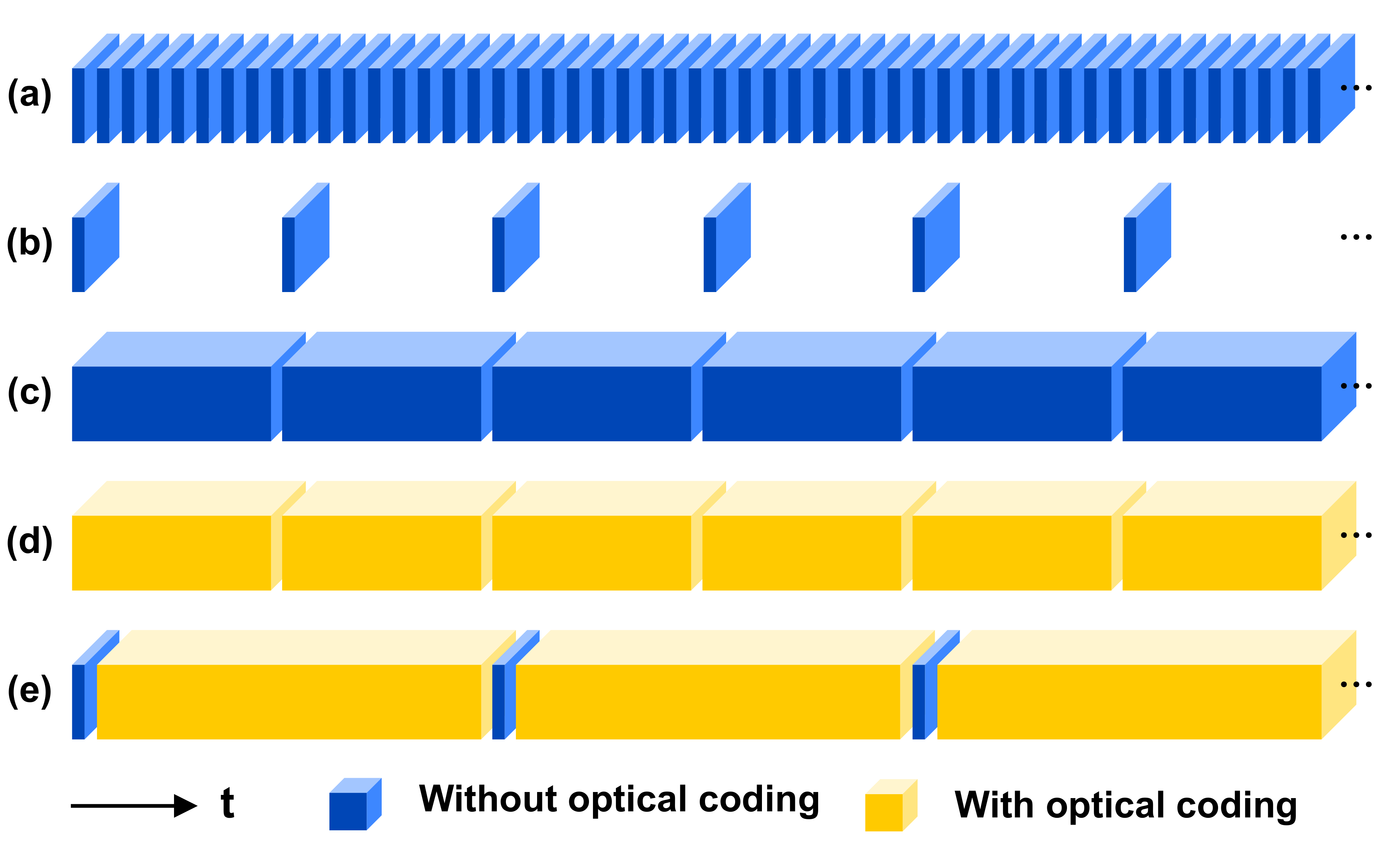}
\end{center}
\caption{Illustration of different video sampling schemes. (a) High frame rate, short exposure. The dynamics can be well recorded, but results in large data size. (b) Low frame-rate, short exposure. Events between two frames will be missed. (c) Low frame rate, long exposure. Motion during exposure will cause blur. (d) Low frame-rate, long exposure with optical coding (standard SCI). Dynamics can be recovered, but with limited visual quality. (e) Proposed KH-CVS, which takes advantage of both short-exposured and long-exposure coded frames.}
\label{fig:scheme_idea}
\end{figure}

Sensing of high-speed scene is widely desirable in applications of robotics, autonomous driving, scientific research, etc. As illustrated in Fig. \ref{fig:scheme_idea}(a), conventional high-speed cameras directly capture such scenes at a high frame rate, while imposing challenges on hardware manufacturing, data readout, transferring, processing, and storage, resulting in massive data sizes and high costs of use. With a low frame rate, off-the-shelf ordinary cameras suffer from loss of interframe information with short exposure [Fig. \ref{fig:scheme_idea}(b)] or motion blur with long exposure [Fig. \ref{fig:scheme_idea}(c)] when capture fast events.

Different from the conventional imaging systems with brute-force sampling strategies, computational imaging \cite{mait2018computational} jointly develops the opto-electronic hardware and the back-end algorithms to realize novel imaging properties. As an important branch of computational imaging, inspired by compressive sensing (CS) theory \cite{donoho2006compressed}, video snapshot compressive imaging (SCI) \cite{yuan2021snapshot} encodes multiple frames into a single snapshot and then computationally makes reconstruction to enable high-speed scene recording with a low-frame-rate sensor [Fig. \ref{fig:scheme_idea}(d)]. To achieve optimal video quality, innovations in both hardware encoder and software decoder techniques have been made in recent years.

Standard video SCI systems \cite{yuan2021snapshot} utilize a series of temporally varying masks to modulate dynamic scene frames at different times, then operate temporal integration through the exposure of the imaging sensor to form a snapshot measurement, namely, a compressive frame. Early video SCI systems \cite{llull2013coded,koller2015high} mechanically translate a static lithography mask to generate temporally varying coding, which faces the challenges of inaccuracy or instability. Currently, mainstream SCI systems use spatial light modulators (SLM) for encoding, such as digital micromirror device (DMD) \cite{qiao2020snapshot,lu2021dual,qiao2020deep} and liquid crystal on silicon (LCoS) \cite{zhang2021ten,zhang2022end}, which enable accurate and flexible pixel-wise exposure control. Recently, extensions to the basic SCI scheme have been developed to enhance the encoding processing. To extend the field-of-view (FoV), the frames of two FoVs are compressed into a snapshot by polarization multiplexing \cite{qiao2020snapshot,lu2021dual}. To improve the spatial resolution of the coding video, the SLM and the static mask can be cascaded to provide finer encoding patterns \cite{zhang2021ten}. Taking advances in deep optics \cite{wetzstein2020inference}, diffractive optical elements (DOEs) have been inserted into the optical path to form an optimized point spread function (PSF) and thus realize spatial superresolution in SCI \cite{zhang2022end}. 

In parallel with hardware systems, research in the field of SCI has also dived deep into the design of reconstruction algorithms. Conventional algorithms employ regularization-based optimization to solve the ill-posed inversion problem. Numerous regularization items, such as total variant (TV) \cite{yuan2016generalized}, non-local low rank \cite{liu2018rank}, Gaussian mixture model (GMM) \cite{yang2014video,yang2014compressive}, and deep denoisers in plug-and-play (PnP) approaches \cite{yuan2020plug,yuan2021plug} serve as priors in optimization frameworks like the alternating direction method of multipliers (ADMM) \cite{boyd2011distributed} or generalized alternating projection (GAP) \cite{liao2014generalized}. These methods are training-free and flexible to different coding but highly time-consuming with hundreds of iterations. Recently, deep neural networks are employed to learning an end-to-end (E2E) mapping from the encoded measurement to the reconstruction video and various network architectures have been proposed. E2E-CNN \cite{qiao2020deep} uses a fully convolutional network with skip-connections, which signiﬁcantly reduces the reconstruction time. BIRNAT \cite{cheng2020birnat} reconstructs the video frames with bidirectional recurrent neural networks (RNN). RevSCI \cite{cheng2021memory} based on multi-group reversible 3D convolutional neural networks has led to state-of-the-art reconstruction results. Taking advantage of advances in the deep unfolding framework, ADMM-net \cite{ma2019deep} unrolls the ADMM iteration loop into an E2E network with a few stages with fast speed and interpretability.

Although sophisticated hardware systems and novel algorithms have been developed to make remarkable progress in SCI, it is still challenging to further improve the visual quality of the reconstructed video. Regardless of implementation differences, they follow the same basic SCI paradigm of extracting multiple frames from a fully blurred compressive frame. The process of video reconstruction is an under-determined inverse problem with inherently high ill-posedness, where a theoretical upper bound of performance is imposed for current paradigm \cite{jalali2019snapshot}. To some extent, the lack of information can be compensated by the priors/regulators or the implicit information contained in trained deep models, making it possible to recover the crude scene appearance. Nonetheless, the recovery of fine details or complicated texture is still arduous. Our insight is that, although lacking dynamic information, short-exposure images can provide a wealth of spatial details, which act conceptually akin to key frames in digital video compression \cite{zhang1995video}. Taking consideration of the temporal consistency of natural scenes, we can alternately sample compressive frames and short-exposure key frames in a hybrid manner and then fuse the two to take both merits for photorealistic reconstruction.

In the field of computational imaging and machine vision, the concept of data fusion is widely established to exploit the complementary properties of multiple sensing modalities. For example, depth data such as steroid camera images, LiDAR point clouds, and radar signals can be fused to enhance the 3D perception \cite{feng2020deep}. In hyperspectral imaging, by fusion of low-resolution hyperspectral images (LR-HSI) with high-resolution multispectral images (HR-MSI) \cite{huang2022deep}, or RGB images with the coded aperture snapshot spectral imaging (CASSI) system \cite{yuan2015side,he2021fast}, the spectral sensing capability can be significantly improved. The event flows provided by a dynamic vision sensors (DVS) can help to ease the motion blur in the images captured by ordinary cameras \cite{pan2019bringing}. Cameras with different spatial-temporal properties can also cooperate to synthesize high-resolution videos \cite{yuan2018temporal,cheng2021dual,paliwal2020deep}. Images captured with different exposure setups are fusion to enhanced the performance of low-light imaging and dynamic range \cite{reinhard2010high}. Under the topic of compressive video sensing, SCI systems can cooperate with long-exposure blurred images captured by standard cameras without coding, to serve as side information (SI) for video reconstruction \cite{yuan2017compressive} or jointly reconstruct high-speed stereo video \cite{sun2017compressive}, but extracting fine details from two blur measurements remains challenging. Furthermore, geometric calibration of cameras is necessary in such scheme, and the data size and expenses are doubled as one more camera is added. Fortunately, the SLMs utilized in currently mainstream SCI systems provide flexible pixel-wise shutter control, enabling compressive frames and short-exposure key frames to be acquired without the need for additional hardware.

In this paper, we propose a new hybrid sampling paradigm for compressive video sensing, dubbed KH-CVS. We alternatively samples short-exposure key frames without coding and long-exposure compressive encoded frames to jointly reconstruct high-speed video. To fuse the hybrid-sampled data, a deep-learning-based reconstruction model is developed. Especially, the optical flow information is leveraged to deal with spatial movements in the dynamic scene. We built a hardware prototype to enable the hybrid sampling scheme. Validated by both numerical evaluation and hardware experiments, the proposed KH-CVS system has been demonstrated superior to conventional snapshot-based methods for photorealistic compressive video sensing.

\section{Methods}
\subsection{Hardware Experiments}
\begin{figure*}[htb]
\begin{center}
\includegraphics[width=1\linewidth]{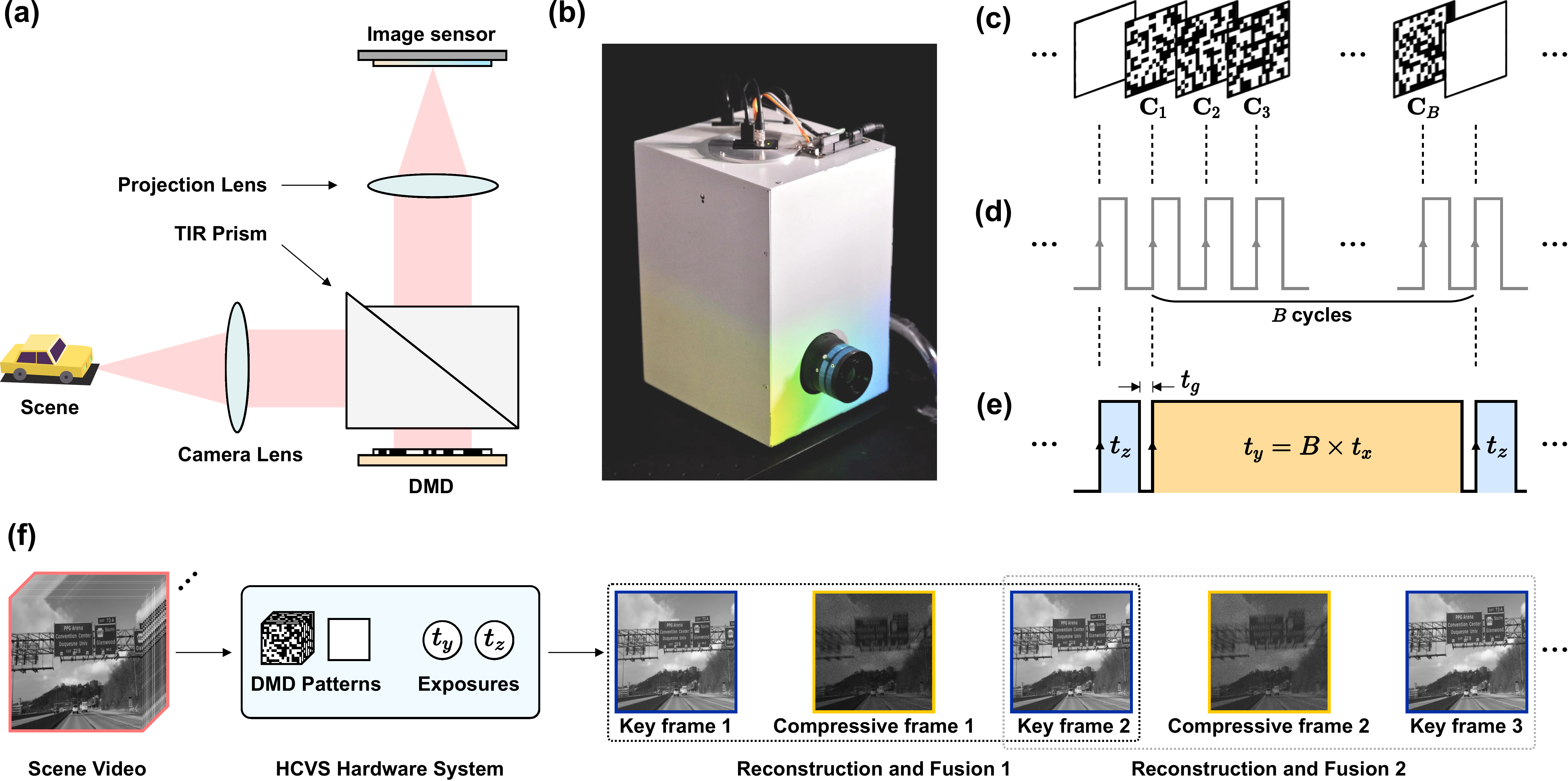}
\end{center}
\caption{Overview of the KH-CVS hardware system. (a) Schematic of the hardware system. DMD: digital micromirror devic; TIR: total internal reflection. (b) A photograph of the prototype built for the KH-CVS experiments. (c)-(d) Conceptual illustration of the optical encoding and exposure scheme. At each rising edge in (d), the DMD updates the pattern displayed on it as (c). The colored area in (e) illustrates the effective exposure of the image sensor. (f) Overview of the sampling scheme, video reconstruction and fusion pipeline in KH-CVS.}
\label{fig:scheme_exp}
\end{figure*}

The optical setup of the KH-CVS prototype built by us is shown in Fig. \ref{fig:scheme_exp}(a). A camera lens forms an image at the focal plane, where a DMD (ViALUX V-9001, 2560 × 1600 resolution, 7.6 um pitch size) is located. After modulated by the DMD, the image is then transferred to an image sensor (FLIR GS3-U3-123S6M-C, 4096 × 3000 resolution, 3.45 um pixel size) by a projection lens with a magnification rate of $0.45\times$ for a pixel-to-pixel mapping. Through the exposure of the image sensor, the measurements can be yielded and readout. The micromirrors on the DMD can be individually rotated $\pm 12^{\circ}$ to reflect the light to different directions, leading to an on-off state control. A total internal reflection (TIR) prism \cite{niu2021fast} is used to steerer the light from the camera lens, making it incident onto the DMD surface at $24^{\circ}$ to meet the angle condition of the DMD. Since it is difficult to directly measure the mismatch between the DMD and the image sensor, a Moire-fringe-based approach proposed by \cite{ri2006accurate} is utilized to align these two elements. In the practice, to fit the spatial resolution of 256 $\times$ 256 in the deep model, we perform 6 $\times$ 6 binning, i.e., capture an region of interest (ROI) of 1536 $\times$ 1536 and then manually downsample by a factor of 6. In order to exclude the ambient light and increase the stability of the system, we package the optical system to build a prototype, of which a photograph is shown in Fig. \ref{fig:scheme_exp}(b).

Since KH-CVS needs to alternately capture short-exposure key frames without coding and long-exposure compressive frames with coding, it is required to design a timing scheme that differs from that used in standard SCI systems. As illustrated in Fig. \ref{fig:scheme_exp}(c), among the preloaded patterns on the DMD, there are $B$ coding patterns ($\mathbf{C}_1$ to $\mathbf{C}_B$) and an additional fully open pattern that all micromirrors are "on" (the DMD works as an ordinary reflective mirror with this pattern). We generate and synchronize the DMD and image sensor control signals through an arbitrary waveform generator (AWG, UNI-T UTG2000B). For the DMD, its trigger signal is a periodic rectangular wave with its period identical to the equivalent exposure time $t_x$, and the DMD refreshes the pattern at each rising edge [Fig. \ref{fig:scheme_exp}(d)]. For the image sensor, the trigger signal is an alternating short and long rectangular wave [Fig. \ref{fig:scheme_exp}(e)]. The short pulse width is $t_x$ and corresponds to a key frame. The duration of the long rectangular wave is $t_y$, and $B$ coding patterns will be consecutively refreshed on the DMD within this exposure, producing a compressive frame. In this manner, a sequence of alternate key frames and compressive frames is captured. A small temporal interval $t_g$ is set between two exposures of the image sensor for data readout and sensor cleaning. 

The conceptual pipeline of KH-CVS is depicted in Fig. \ref{fig:scheme_exp}(f). With the aforementioned setup of DMD patterns and trigger signals, the scene video with high temporal dimension is encoded and captured as a sequence of key frames and compressive frames. Similar to standard SCI, $B$ frames can be recovered from each compressive, which is then fused with its neighboring two key frames, as indicated by the black dotted box. For compressive frame 1, key frame 1 and 2 are referred to as left and right key frame. Key frame 2 is also reused as the left key frame for compressive frame 2 in the gray dotted box. Thus, averaged 2 frames are capable of producing $B+1$ frames ($B$ reconstructed frames and 1 key frames itself) in KH-CVS, which leads to a compressive ratio as $2/(B+1)$ from a systematic viewpoint.

\begin{figure}[htbp]
\begin{center}
\includegraphics[width=0.6\linewidth]{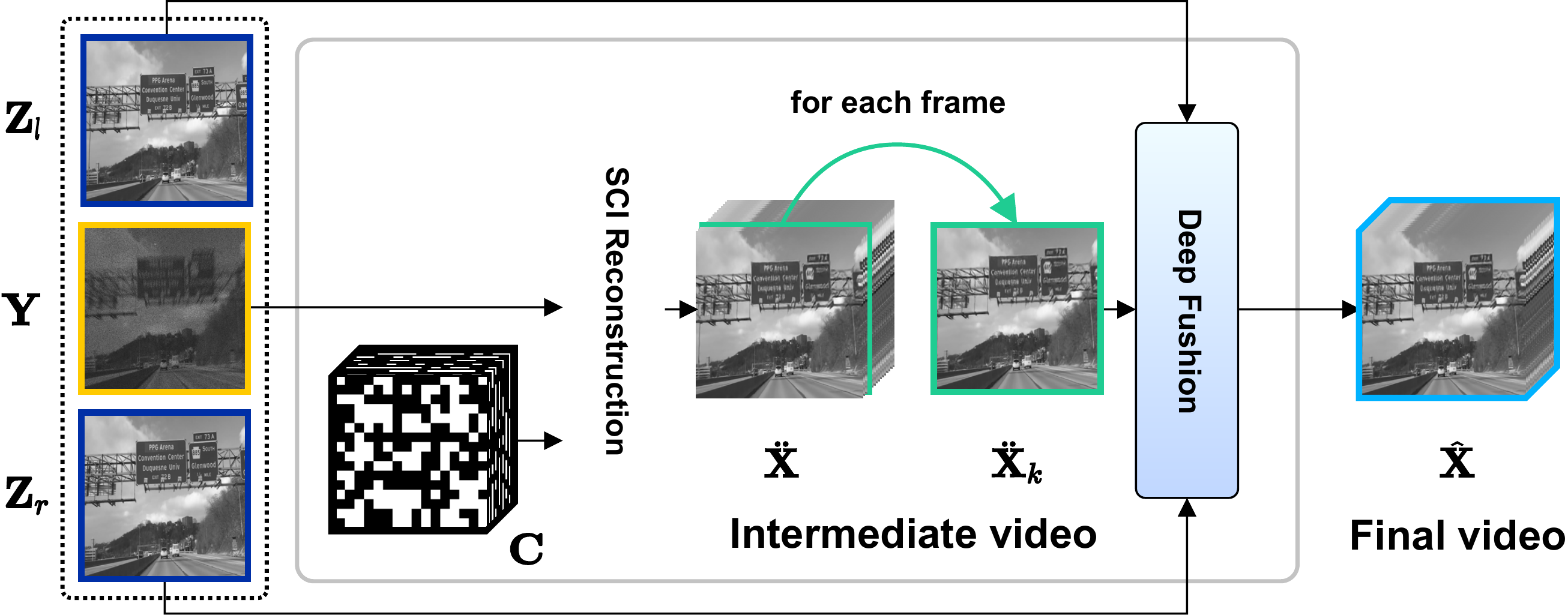}
\end{center}
\caption{Illustration of the reconstruction and fusion pipeline. The basic SCI algorithm reconstructs an intermediate video, then each frame of it is fed into a deep fusion module with the two key frames to yield final outputs.}
\label{fig:scheme_main_b}
\end{figure}

\begin{figure*}[htbp]
\begin{center}
\includegraphics[width=1\linewidth]{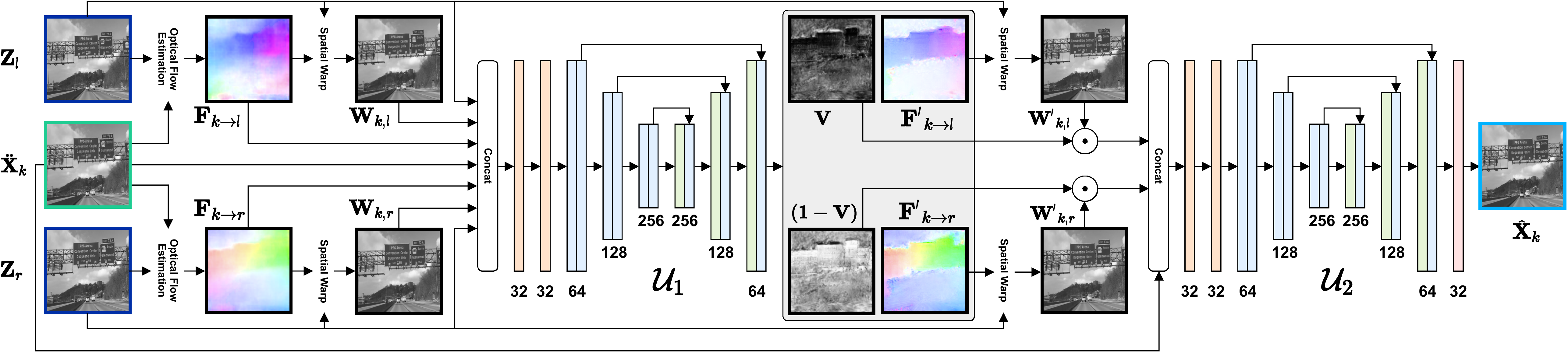}
\end{center}
\caption{Schematic of the fusion module. The numbers below the layers indicate feature map channels and $\odot$ denotes the Hadamard product.}
\label{fig:scheme_main}
\end{figure*}

As illustrated in Fig. \ref{fig:scheme_main_b}, in the algorithm, an intermediate reconstructed video is first made with a standard SCI method, and then each frame of it is fused with the two key frames by a deep fusion module (Fig. \ref{fig:scheme_main}). The detailed methods are introduced as below.

\subsection{Standard Video SCI}
In the mathematical model of standard video SCI, a high-speed dynamic scene video $\mathbf{X} \in \mathbb{R}^{n_x \times n_y \times B}$ containing $B$ frames $\left\{\mathbf{X}_{k}\right\}_{k=1}^{B} \in \mathbb{R}^{n_{x} \times n_{y}}$ with spatial resolution $n_{x} \times n_{y}$, is modulated by a coding cube $\mathbf{C} \in \mathbb{R}^{n_x \times n_y \times B}$, which contains $B$ encoding masks $\left\{\mathbf{C}_{k}\right\}_{k=1}^{B} \in \mathbb{R}^{n_{x} \times n_{y}}$. In general, pre-generated pseudo-random patterns are established for encoding in SCI systems. During the long exposure time $t_y$ of the image sensor, the coded frames are temporally integrated to produce a coded 
compressive frame $\mathbf{Y} \in \mathbb{R}^{n_x \times n_y}$, which is 
\begin{equation}
\mathbf{Y} = \sum_{k=1}^{B} \mathbf{C}_k \odot \mathbf{X}_k+\mathbf{G},
\end{equation}
where $\mathbf{G} \in \mathbb{R}^{n_x \times n_y}$ represents the measurement noise and $\odot$ denotes the Hadamard product, which can be expressed in detail as 
\begin{equation}
y_{ij}=\sum_{k=1}^{B} c_{ijk} x_{ijk}+g_{ij},
\end{equation}
where $i \in [1,2, \ldots, n_x]$ and $j \in [1,2, \ldots, n_y]$ denote spatial positions and $y$, $c$, $x$, $g$ are elements of $\mathbf{Y}$, $\mathbf{C}$, $\mathbf{X}$, $\mathbf{G}$, respectively. By taking the prior-known encoding masks $\mathbf{C}$ and the compressive frame $\mathbf{Y}$ as inputs, a reconstruction algorithm for standard SCI $\mathcal{R}$ is applied to estimate the scene video $\ddot {\mathbf{X}} \in \mathbb{R}^{n_x \times n_y \times B}$, which contains $B$ reconstructed frames $\left\{\ddot {\mathbf{X}}_{k}\right\}_{k=1}^{B} \in \mathbb{R}^{n_{x} \times n_{y}}$:
\begin{equation}
\ddot {\mathbf{X}} = \mathcal{R}(\mathbf{Y},\mathbf{C}),
\end{equation}
and the equivalent exposure time $t_x$ of each frame of the reconstruction video is $t_y/B$, thus the equivalent frame rate is boosted by $B$ times. For standard SCI, $\ddot {\mathbf{X}}$ is the output as the final results, but in KH-CVS, it acts as an intermediate reconstruction, which is then fused with the key frames to achieve enhanced visual quality.

\begin{figure*}[htb!]
\begin{center}
\includegraphics[width=1\linewidth]{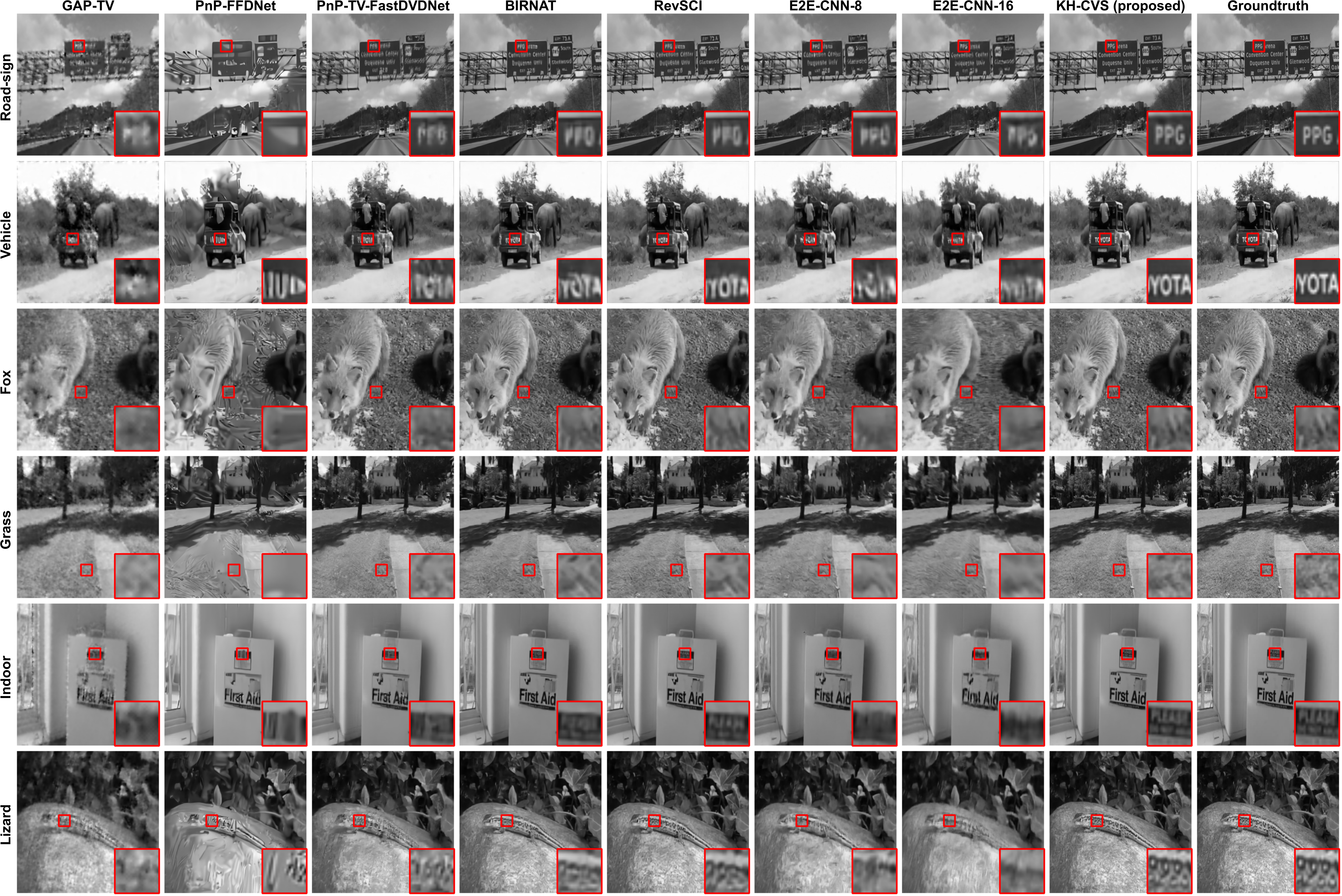}
\end{center}
\caption{Reconstructed frames of GAP-TV, PnP-FFDNet, PnP-TV-FastDVDNet, BIRNAT, RevSCI (all with $B=8$), E2E-CNN with $B=8$ and $B=16$, and proposed KH-CVS with $B=16$ on six simulated video datasets. The details in red boxes are enlarged for better visualization. Please see Visualization 1 for the reconstructed videos.}
\label{fig:sim_main}
\end{figure*}

\subsection{Hybrid Sampling and Deep Fusion}
In the proposed KH-CVS, the short-exposure key frames and the long-exposure encoded frames are alternatively sampled, as shown in Fig. \ref{fig:scheme_exp}(f). For each compressive frame $\mathbf{Y}$, its two neighboring short-exposured key frames $\mathbf{Z}_l,\mathbf{Z}_l \in \mathbb{R}^{n_{x} \times n_{y}}$ are 
\begin{equation}
\mathbf{Z}_l = \mathbf{X}_{0}+\mathbf{G},\quad\mathbf{Z}_r = \mathbf{X}_{B+1}+\mathbf{G}.
\end{equation}

To match the equivalent exposure time $t_x$, the exposure time of the key frames is set to $t_z=t_x$. As illustrated in Fig. \ref{fig:scheme_main_b}, each frame $\ddot{\mathbf{X}}_k$ in $\ddot {\mathbf{X}}$ is sequentially fused with $\mathbf{Z}_l$ and $\mathbf{Z}_r$ to produce the corresponding output frame $\hat{\mathbf{X}}_k$ by the fusion module $\mathcal{H}$ shown in Fig. \ref{fig:scheme_main}:
\begin{equation}
\hat {\mathbf{X}}_k = \mathcal{H}(\mathbf{Z}_l,\mathbf{Z}_r,\ddot {\mathbf{X}}_k), \ \text{for} \  k = 1, 2, \ldots, B.
\end{equation}

Although the key frames contain delicate visual details, there are movements between them and the intermediate frame, causing the features or contents of them to be spatially misaligned, posing a major difficulty for image fusion. To overcome this problem, we incorporate the optical flow \cite{baker2011database} into the model, which is able to reveal the moving velocity and direction of each pixel between two frames. As a widely used technique, optical flow can be calculated with an off-the-shelf estimator, noted as $\mathcal{F}$. Then, the key frames can be spatially warped to align with the target intermediate frame for further fusion. To this end, the first step is to estimate the optical flow from the $k$th intermediate frame $\ddot{\mathbf{X}}_k$ to $\mathbf{Z}_l$ and $\mathbf{Z}_r$,
\begin{equation}
\mathbf{F}_{k \to l} = \mathcal{F}(\ddot{\mathbf{X}}_k,\mathbf{Z}_l),\quad\mathbf{F}_{k \to r} = \mathcal{F}(\ddot{\mathbf{X}}_k,\mathbf{Z}_r).
\end{equation}
Then using a spatial warping function $\mathcal{W}$, which is an interpolation operation \cite{baker2011database} based on the optical flow, the two key frames are warped to the target frame as 
\begin{equation}
\mathbf{W}_{k,l}=\mathcal{W}(\mathbf{Z}_l,\mathbf{F}_{k \to l}),\quad\mathbf{W}_{k,r}=\mathcal{W}(\mathbf{Z}_r,\mathbf{F}_{k \to r}).
\end{equation}

Due to the relatively low quality of intermediate frames, the optical flows and warped frames in this stage are relatively rough. Thus, a neural network $\mathcal{U}_1$ is used to refine optical flows by estimating their residuals. Moreover, intuitively, if the target frame is temporally near to the left keyframe, the contribution of the warped left keyframe to the final fused frame should be weighted more, and vice versa. Thus, a visible map is introduced to weight the pixel values taken from two warped keyframes, which is inspired by \cite{jiang2018super}. We use one output channel of $\mathcal{U}_1$ with an sigmoid function to yield a visible map $\mathbf{V}$ between [0,1] to weight left keyframe and use $(1-\mathbf{V})$ for right keyframe. 

\begin{equation}
\left\{\Delta\mathbf{F}_{k \to l},\Delta\mathbf{F}_{k \to r}, \mathbf{V} \right\} = \mathcal{U}_1(\mathbf{Z}_l,\mathbf{Z}_r,\mathbf{W}_{k,l},\mathbf{W}_{k,r},\ddot{\mathbf{X}}_k).
\end{equation}
Thus, the refined optical flows are calculated by
\begin{equation}
\mathbf{F'}_{k \to l} = \mathbf{F'}_{k \to l} + \Delta\mathbf{F'}_{k \to l},\quad\mathbf{F'}_{k \to r} = \mathbf{F'}_{k \to r} + \Delta\mathbf{F'}_{k \to r},
\end{equation}
which are used to warp the key frames as 
\begin{equation}
\mathbf{W}_{k,l}'=\mathcal{W}(\mathbf{Z}_l,\mathbf{F'}_{k \to l}),\quad\mathbf{W}_{k,r}'=\mathcal{W}(\mathbf{Z}_r,\mathbf{F'}_{k \to r}).
\end{equation}
Finally, another network $\mathcal{U}_2$ is used to make the final output frame by synthesizing the visible maps, warped key frames with optimal optical flows, and the intermediate frame:
\begin{equation}
\hat{\mathbf{X}}_k = \mathcal{U}_2(\mathbf{V}\odot\mathbf{W}_{k,l}',(1-\mathbf{V})\odot\mathbf{W}_{k,r}',\ddot{\mathbf{X}}_k).
\end{equation}

Using the same method, all frames in $\ddot{\mathbf{X}}$ are processed and combined to form the final video $\hat{\mathbf{X}}$.

\subsection{Model implementation and network training}
The KH-CVS is a flexible framework that allows for a variety choices of submodules. For the demonstration in this work, E2E-CNN \cite{qiao2020deep} is adopted as the basic SCI reconstruction model $\mathcal{R}$. For optical flow calculation, taking both efficiency and precision into account, a well-established PWC-Net \cite{sun2018pwc} is used as optical flow estimator $\mathcal{F}$. To improve the accuracy of the optical flow, instead of directly calculate $\mathbf{F}_{k \to l}$, we calculate $\mathbf{F}_{k \to k-1}, \mathbf{F}_{k-1 \to k-2},\dots,\mathbf{F}_{1 \to l}$, and combine the relatively small movements revealed by this set of optical flows to form $\mathbf{F}_{k \to l}$. As illustrated in Fig. \ref{fig:scheme_main}, the networks $\mathcal{U}_1$ and $\mathcal{U}_2$ are based on U-Net \cite{ronneberger2015u}. Inspired by \cite{niklaus2018context,paliwal2020deep}, the shallow layers in CNN are able to extract edges and textures as contextual information to help produce better results. Therefore, we use the first convolutional layer in a ResNet18 \cite{he2016deep} to extract the feature maps of the frames and concatenate them together before feeding them into $\mathcal{U}_2$. In order to improve the ability of the network to handle spatial deformation, at the end of $\mathcal{U}_2$, a deformable convolutional layer \cite{dai2017deformable} is applied to synthesize deep latent features to produce the final output.

The model is trained in a multi-stage manner. First, the SCI reconstruction model $\mathcal{R}$ is trained following the configuration of the corresponding SCI model (which is E2E-CNN here). For the loss function, the L1 distance is adopted here, which is more favorable than L2 distance in image restoration tasks \cite{zhao2016loss}. Thus, the loss function of this stage is 
\begin{equation}
\mathcal{L}_{rec} = || \ddot {\mathbf{X}} - {\mathbf{X}} ||_1.
\end{equation}

Then, the optical flow refinement model $\mathcal{U}_1$ is trained. Given that $\mathcal{U}_2$ is not ready at this stage, we can use the aforementioned visible map to form a linear combination of two warped keyframes to supervise the training of $\mathcal{U}_1$, 
\begin{equation}
\hat{\mathbf{X}}_k' = \frac{(1-\tau)\mathbf{V}\odot\mathbf{W}_{k,l}'+\tau(1-\mathbf{V})\odot\mathbf{W}_{k,r}'}{(1-\tau)\mathbf{V}+\tau(1-\mathbf{V})},
\end{equation}
where $\tau = k/(B+2) \in (0,1)$ is a time factor that indicates the relative temporal distance to the left and right key frames. To supervise both synthesized frames and warped frames, the loss function for training $\mathcal{U}_1$ is
\begin{equation}
\mathcal{L}_{int} = \lambda_1 || \hat{\mathbf{X}}_k' - {\mathbf{X}}_k ||_1 + \lambda_2 ( || \mathbf{W}_{k,l}' - {\mathbf{X}}_k ||_1 + || \mathbf{W}_{k,r}'- {\mathbf{X}}_k ||_1 ).
\end{equation}

Finally, $\mathcal{U}_2$ is trained using the loss calculated with the final output as
\begin{equation}
\mathcal{L}_{out} = || \hat {\mathbf{X}} - {\mathbf{X}} ||_1.
\end{equation}

The proposed architecture is implemented in PyTorch with two NVIDIA RTX 3090 GPUs. Adam optimizer \cite{kingma2014adam} is used to train the model, with its parameters $\beta_1$ and $\beta_2$ set to 0.5 and 0.999. For training of the basic SCI model $\mathcal{R}$, the learning rate is set to $10^{-4}$ for 100 epochs and then $10^{-5}$ for another 100 epochs. Then the trained $\mathcal{R}$ is fixed and $\mathcal{U}_1$ is trained with the same setup of learning rate and epoch number, with $\lambda_1$ and $\lambda_2$ set to 200 and 100, respectively. Next, $\mathcal{U}_2$ is connected to the model and trained 50 epochs with an initial learning rate of $10^{-5}$ and then $10^{-6}$ for another 50 epochs. Finally, the whole model is jointly trained for 50 epochs at a learning rate of $10^{-6}$. For model training, 5000 video clips are sampled from the ImageNet VID \cite{russakovsky2015imagenet} training set and randomly cropped to 256 $\times$ 256, with horizontal flipping randomly for data augmentation. Each video clip has 16 continuous frames (thus $B=16$) with their two neighbored key frames to serve as network inputs. 

\section{Experiments}
\subsection{Simulation Experiments}
\renewcommand\arraystretch{1.1}
\begin{table*}[htb!]
\caption{Quantitative comparison with other methods, measured by average PSNR (in dB), SSIM, and LPIPS. The higher PSNR and SSIM scores, and lower LPIPS scores, the better quality of reconstructions. The numbers in \textbf{bold} represent the best performance.}
\begin{center}
\resizebox{1\textwidth}{!}{
\begin{tabular}{ccccccccc}
\hline
Metrics                & Methods            & Road-sign & Vehicle & Fox   & Grass  & Indoor & Lizard & \textit{Average} \\
\hline
\multirow{8}{*}{PSNR}  & GAP-TV             & 22.94    & 24.75   & 23.52  & 24.27  & 28.22  & 24.31  & 24.67   \\
                       & PnP-FFDNet         & 22.07    & 24.10   & 22.58  & 24.27  & 30.69  & 23.92  & 24.61   \\
                       & PnP-TV-FastDVDNet  & 26.42    & 28.36   & 26.44  & 28.09  & 34.98  & 26.17  & 28.41   \\
                       & BIRNAT             & 29.32    & 28.77   & 26.73  & 28.81  & 37.50  & 29.78  & 30.15   \\
                       & RevSCI             & 29.78    & \textbf{30.30}   & 28.47  & 29.56  & \textbf{38.11}	& 30.77  & 31.16   \\
                       & E2E-CNN-8          & 28.63    & 28.49   & 25.51  & 27.52  & 34.58  & 27.88  & 28.77   \\
                       & E2E-CNN-16         & 27.45    & 26.84   & 24.19  & 26.12  & 31.63  & 26.34  & 27.10   \\
                       & KH-CVS (proposed)   & \textbf{31.62}    & 29.79   & \textbf{29.84}  & \textbf{32.14}  & 37.27  & \textbf{32.58}  & \textbf{32.21}   \\
\hline
\multirow{8}{*}{SSIM}  & GAP-TV             & 0.7885   & 0.7938  & 0.6193 & 0.6544 & 0.8842 & 0.6817 & 0.7370  \\
                       & PnP-FFDNet         & 0.7884   & 0.7975  & 0.5940 & 0.6634 & 0.9441 & 0.6730 & 0.7434  \\
                       & PnP-TV-FastDVDNet  & 0.9090   & 0.8991  & 0.7988 & 0.8120 & 0.9691 & 0.7906 & 0.8631  \\
                       & BIRNAT             & 0.9520   & 0.9127  & 0.8203 & 0.8410 & 0.9817 & 0.8992 & 0.9012  \\
                       & RevSCI             & 0.9560   & \textbf{0.9356}  & 0.8801 & 0.8644 & \textbf{0.9837} & 0.9213 & 0.9235  \\
                       & E2E-CNN-8          & 0.9459   & 0.9036  & 0.7383 & 0.7925 & 0.9709 & 0.8374 & 0.8648  \\
                       & E2E-CNN-16         & 0.9254   & 0.8537  & 0.6227 & 0.7241 & 0.9464 & 0.7607 & 0.8055  \\
                       & KH-CVS (proposed)    & \textbf{0.9689}	 & 0.9279  & \textbf{0.9175}	& \textbf{0.9416}	& 0.9808	& \textbf{0.9460}	& \textbf{0.9471}  \\
\hline
\multirow{8}{*}{LPIPS} & GAP-TV             & 0.2934   & 0.2671  & 0.3507 & 0.3945 & 0.2314 & 0.3001 & 0.3062  \\
                       & PnP-FFDNet         & 0.2802   & 0.2651  & 0.4322 & 0.4005 & 0.0716 & 0.3387 & 0.2980  \\
                       & PnP-TV-FastDVDNet  & 0.1028   & 0.1010  & 0.1551 & 0.1591 & 0.0391 & 0.1574 & 0.1191  \\
                       & BIRNAT             & 0.0598   & 0.0806  & 0.1259 & 0.1556 & 0.0180 & 0.0745 & 0.0857  \\
                       & RevSCI             & 0.0492   & 0.0567  & 0.1065 & 0.1328 & \textbf{0.0129} & 0.0547 & 0.0688  \\
                       & E2E-CNN-8          & 0.0752   & 0.1214  & 0.2412 & 0.2491 & 0.0404 & 0.1461 & 0.1456  \\
                       & E2E-CNN-16         & 0.1120   & 0.1942  & 0.3487 & 0.3517 & 0.0892 & 0.2068 & 0.2171  \\
                       & KH-CVS (proposed)    & \textbf{0.0316}	 & \textbf{0.0465} & \textbf{0.0518} & \textbf{0.0446}	& 0.0195	& \textbf{0.0380}	& \textbf{0.0387}  \\
\hline
\end{tabular}}
\end{center}
\label{table:sim_metrics}
\end{table*}

A series of simulation experiments are conducted to investigate the performance of the proposed KH-CVS on simulated datasets from ImageNet VID \cite{russakovsky2015imagenet} test set and NFS \cite{kiani2017need}, including Road-sign, Vehicle, Fox, Grass, Indoor and Lizard. Multiple existing algorithms are used for comparison, including GAP-TV \cite{yuan2016generalized}, PnP-FFDNet \cite{yuan2020plug}, PnP-TV-FastDVDNet \cite{zhang2021ten}, BIRNAT \cite{cheng2020birnat} and RevSCI \cite{cheng2021memory}. According to the analysis in Sec.2, the compressive ratio is $2/(B+1)$, i.e., 11.67\% for $B=16$. In order to conduct a fair comparison, the systematic compressive ratio of the competitive counterpart algorithms should not be lower than KH-CVS, ensuring sufficient data amount for the counterparts. Therefore, for the counterparts, the same 16 frames are reconstructed from two snapshot (each encodes 8 frames in standard SCI paradigm), which corresponds to a compressive ratio of 12.5\%. As the basic SCI model used in KH-CVS, the E2E-CNN \cite{qiao2020deep} is also compared. E2E-CNN-8 and E2E-CNN-16 represent the models trained with $B=8$ and $B=16$, respectively. 

Fig. \ref{fig:sim_main} shows the exemplary frames of the reconstructed videos, which validates that KH-CVS is capable of providing higher visual quality with more fine details. The baseline models GAP-TV results suffer from strong noise and the PnP-FFDNet produces unpleasant artifacts. Among the countpart methods, the state-of-the-art algorithm RevSCI shows the best performance, yet the details in the reconstruction are relatively blurry and distorted (like the characters in scene Road-sign and Vehicle). In the basic E2E-CNN-16 model of KH-CVS, Details are also blurred out, but rough outlines are reserved. By fusion of the information from the key frames, the details can be well compensated to obtain high-quality reconstructions.

To quantitatively evaluate reconstruction quality, multiple metrics are used, including widely-used peak signal to noise ratio (PSNR) \cite{wang2004image}, structural similarity index measure (SSIM) \cite{wang2004image}. In addition to these classic metrics, the learned perceptual image patch similarity (LPIPS) \cite{zhang2018unreasonable} is also used to evaluate perceptual visual quality, which is calculated in the deep feature space as perceptual distance. Note that a lower LPIPS means the reconstructions are more similar to the ground-truths. Since the key frames are identical to their corresponding ground-truth frames in the simulation, metrics are calculated only on the reconstructed frames to prevent privilege of KH-CVS. Quantitative results are summarized in Tab. \ref{table:sim_metrics}. It can be observed that the proposed KH-CVS achieves 1.05 dB improvement in PSNR and 0.0236 in SSIM on average, compared to the best results of counterparts, i.e., RevSCI. KH-CVS reduces the averaged LPIPS to about half of RevSCI, which reveals that the KH-CVS reconstruction frames are perceptually more closed to the ground-truth frames. It is also worth mentioning that, compared to its basic SCI model E2E-CNN-16, KH-CVS boosts PSNR by 5.11 dB and SSIM by 0.1416, and decreases LPIPS to about only 17.8\% of itself, which shows that the information from the key frames can help significantly improve the performance of the basic SCI model.

\subsection{Visualization of Intermediate Results in the Pipeline}
\begin{figure*}[htb]
\begin{center}
\includegraphics[width=1\linewidth]{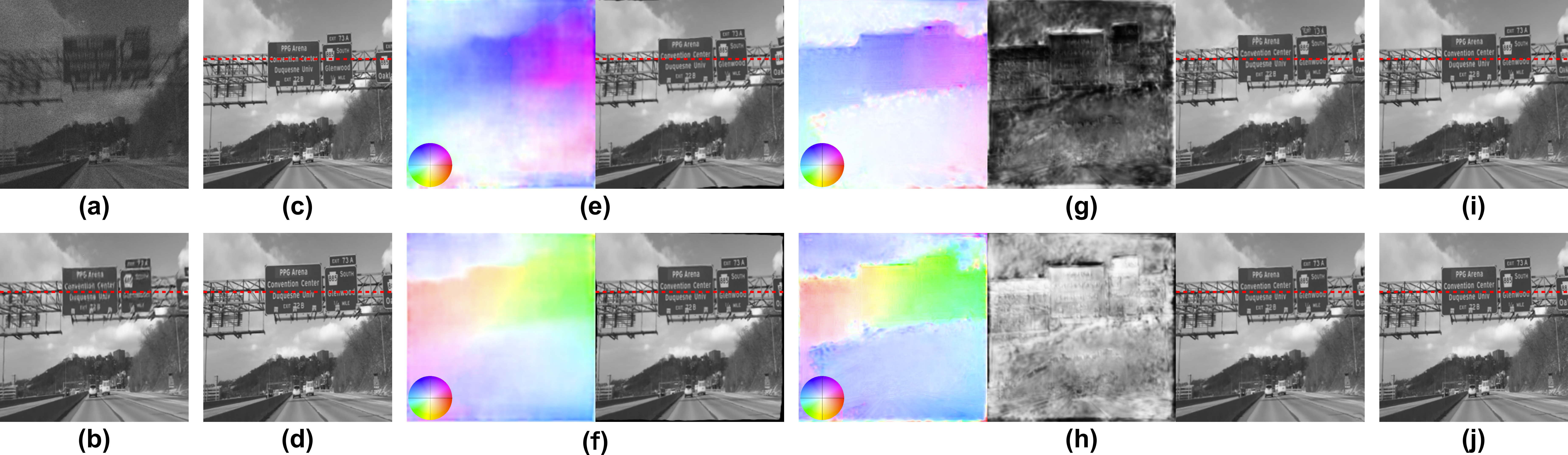}
\end{center}
\caption{Visualization of the intermediate results in the KH-CVS pipeline. (a) The coded compressive frame $\mathbf{Y}$. (b) An exemplary frame $\ddot {\mathbf{X}}_{12}$ from the intermediate reconstruction $\ddot {\mathbf{X}}$. (c) Left key frame $\mathbf{Z}_l$. (d) Right key frame $\mathbf{Z}_r$. (e) Original left optical flow $\mathbf{F}_{12 \to l}$ and the corresponding warped left key frame $\mathbf{W}_{12,l}$. (f) Original right optical flow $\mathbf{F}_{12 \to r}$ and the corresponding warped left key frame $\mathbf{W}_{12,r}$. (g) Refined left optical flow $\mathbf{F'}_{12 \to l}$, visible map $\mathbf{V}$, and the corresponding warped left key frame $\mathbf{W}_{12,l}'$. (h) Refined right optical flow $\mathbf{F'}_{12 \to r}$, visible map $(1-\mathbf{V})$, and corresponding warped right key frame $\mathbf{W}_{12,r}'$. (i) Output frame after fusion $\hat {\mathbf{X}}_{12}$. (j) Ground-truth frame ${\mathbf{X}}_{12}$. The red dash lines serve as references to indicate the motion. The color coding legends are shown with the optical flows.}
\label{fig:sim_vis}
\end{figure*}

The inputs and intermediate results of the model are visualized in Fig. \ref{fig:sim_vis}. Here the 12th frame $\hat {\mathbf{X}}_{12}$ is used as an example. To visualize the optical flow field, we color code the optical flow field following \cite{baker2011database}, that is, using the color hue to represent the direction of motion and the saturation to represent the magnitude. In the scene of Road-sign, as the vehicle moves forward, the road-sign are from far to near, so from the left key frame $\mathbf{Z}_l$ to $\ddot {\mathbf{X}}_{12}$, the road signs are expanded outward, and the preliminarily estimated optical flow $\mathbf{F}_{12 \to l}$ is consistent with this observation. Through this optical flow, $\mathbf{Z}_l$ can be warped to $\ddot {\mathbf{X}}_{12}$ as $\mathbf{W}_{12,l}$. Because $\mathbf{F}_{12 \to l}$ is relatively rough, the warp results $\mathbf{W}_{12,l}$ also have errors (such as the position of the characters and the edge of the image). After refinement, the optical flow $\mathbf{F'}_{12 \to l}$ is more accurate and can distinguish the area of the road-sign with a displacement and the area of the relatively static 
background. At this time, the warped result $\mathbf{W}_{12,l}'$ is also enhanced. The behavior of the right key frame $\mathbf{Z}_r$ is similar to that mentioned above. Since $\mathbf{Z}_r$ is temporally nearer to $\ddot {\mathbf{X}}_{12}$ here, it is supposed to provide more information than $\mathbf{Z}_l$, which is also consistent with that revealed by the visible map $\mathbf{V}$. It can be seen that, compared to the intermediate reconstructed frame $\ddot {\mathbf{X}}_{12}$, the final output frame $\hat {\mathbf{X}}_{12}$ has a significant improvement in the details of the scene, which is closer to the ground-truth ${\mathbf{X}}_{12}$.

\subsection{Validation of the Robustness to the Frame Gap}
\begin{figure}[htb]
\begin{center}
\includegraphics[width=0.6\linewidth]{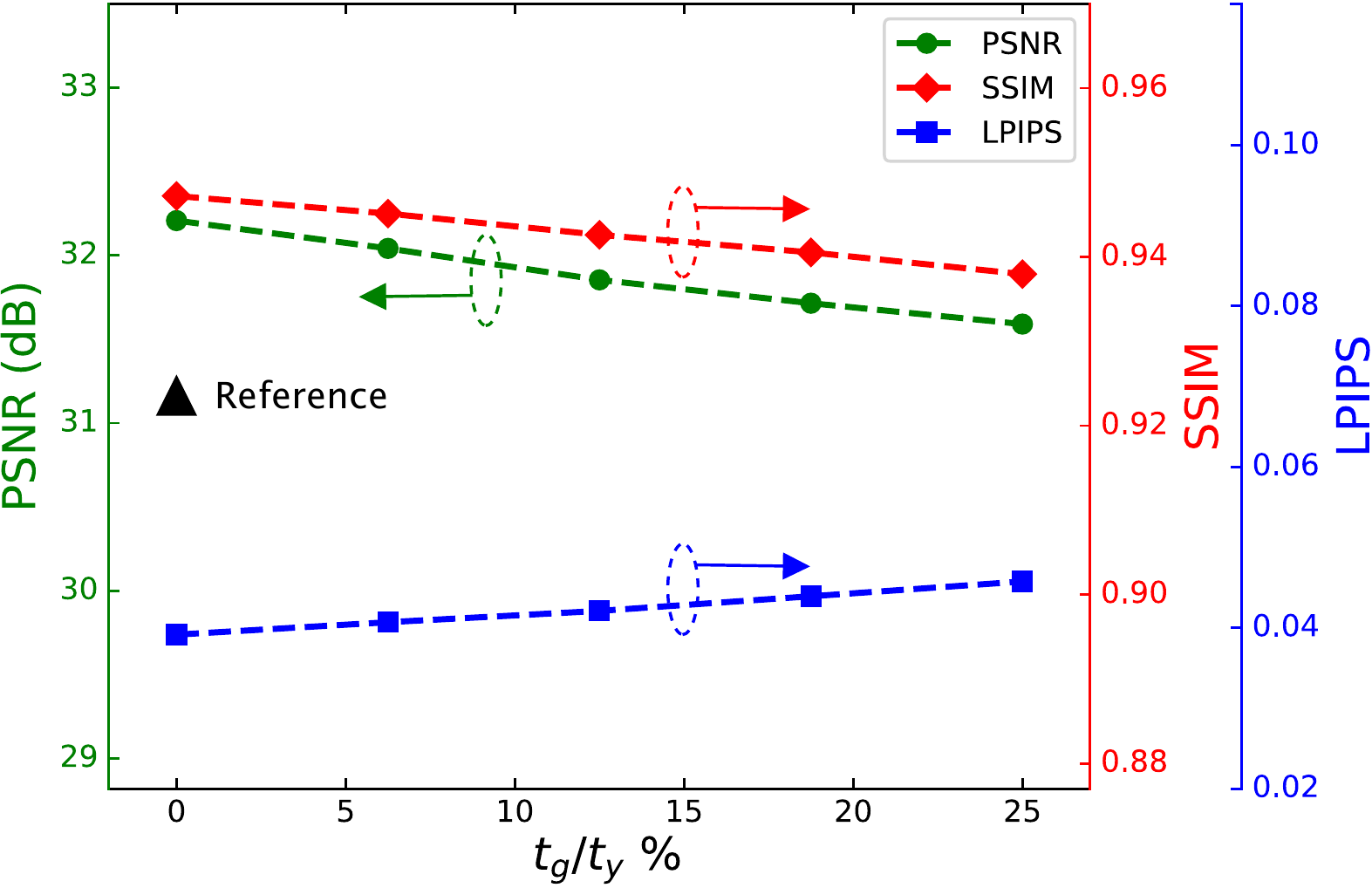}
\end{center}
\caption{Robustness of KH-CVS to the frame gap. PSNR, SSIM and LPIPS comparison on different normalized frame gap $t_g/t_y$ (in percentage). The black triangle stands for the metrics of the best counterpart, i.e., RevSCI, with its location applicable to all three vertical axes.}
\label{fig:frame_gap}
\end{figure}

In the process of capturing, the image sensor needs time to readout the image data and clear the sensor to prepare for next image, thus the frame gap (noted as $t_g$) exists in real systems. This gap also occurs between two snapshot in the standard SCI. Since KH-CVS fuses multiple frames to make reconstruction with considering temporal consistency of the scene, its robustness to the gap between measured frames is needed to be evaluated. In this subsection, a series of experiments on simulated datasets are conducted to evaluate the robustness to the frame gap. In the ideal case, the $B$ frames to be compressed and the two key frames are continuously sampled from the video dataset. In order to simulate the actual frame gap, we skip some frames between the key frames and $B$ compressive frames during this test. For example, when $t_g=2t_x$, it means that there are 2 skipped frames between the left key frame $\mathbf{Z}_l$ and the first compressive frame $\mathbf{X}_0$, and also 2 frames between $\mathbf{Z}_r$ and $\mathbf{X}_B$, thus the normalized frame gap $t_g/t_y=2/16=12\%$. In order to evaluate the effect of frame gap on reconstruction quality, we sweep the frame gap from 0 to 4 frames, corresponding to the $t_g/t_y$ changing from 0 to 25\%, and the results are presented in Fig. \ref{fig:frame_gap}. It can be seen from the results that with the increase of the $t_g/t_y$, the quality of reconstruction decreases, but the trend is relatively slight. From the ideal situation to $t_g/t_y=25\%$, the PSNR drops by 0.61 dB, the SSIM drops by 0.0092, and the LPIPS rises by 0.0066, while all metrics are still better than the best of the counterpart methods (marked by the black triangle). This shows that KH-CVS is robust to the frame gap.  We experimentally found that setting $t_g$ as 300 $\mu s$ is enough for preventing missing the next trigger pulse and allows a continuously sampling. For the demonstration in this paper, $B=16$, $t_y$ is 33328 (see below) $\mu s$, resulting in $t_g/t_y$ less than 1\%. Thus the reconstruction model is capable to deal with the frame gap in this system setup.

\subsection{Hardware Experiments}
\begin{figure*}[htb!]
\begin{center}
\includegraphics[width=1\linewidth]{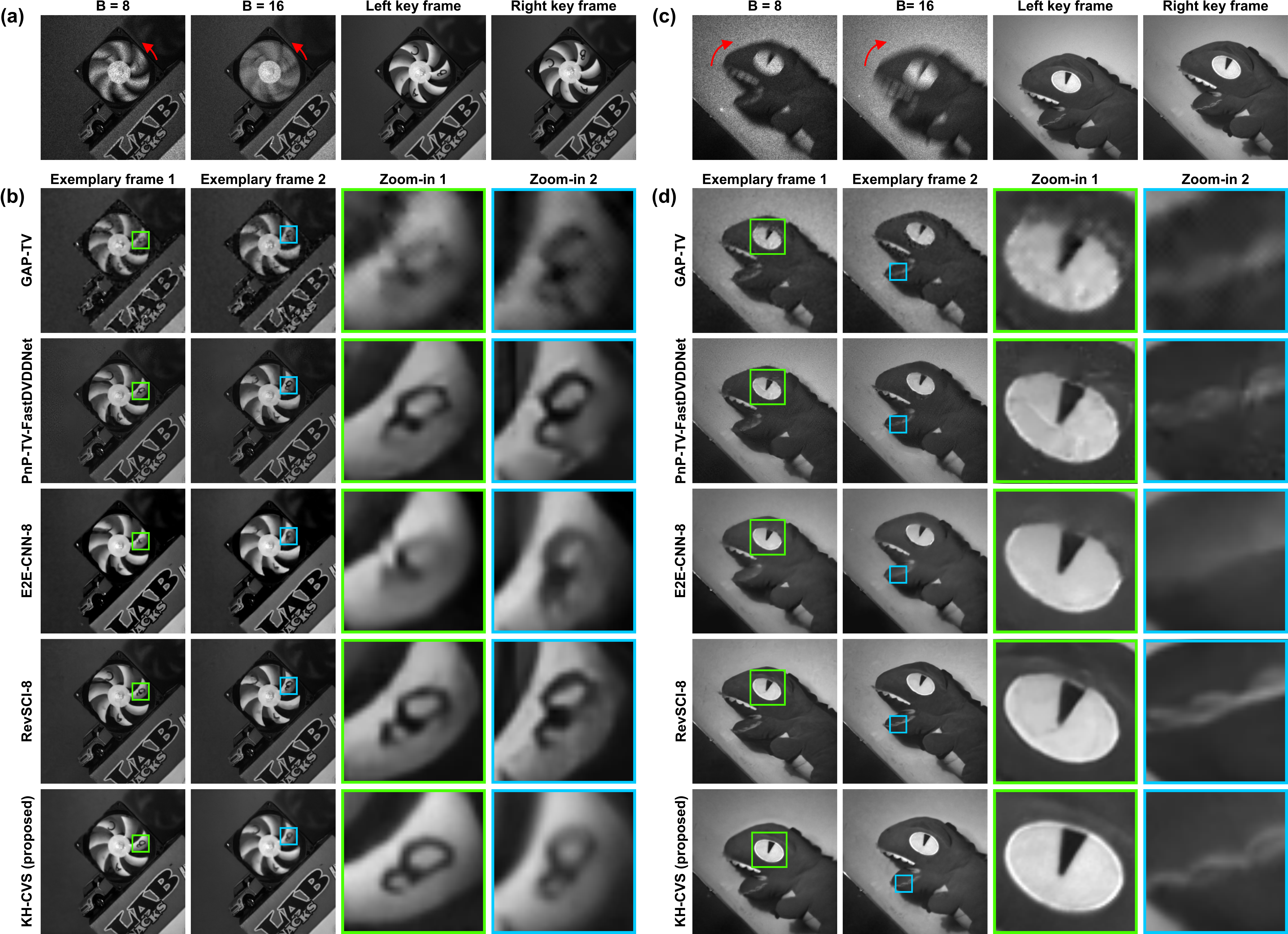}
\end{center}
\caption{Experimental results in natural scenes. (a) and (c) Compressive frames of $B=8$ and $B=16$, left and right key frames, for the rotating fan and the moving toy dinosaur scenes, respectively. The red arrows indicate the motion directions of the objects. (b) and (d) Exemplary frames reconstructed by different methods with corresponding zoom-in views. Results of GAP-TV, PnP-TV-FastDVDNet, E2E-CNN-8 and RevSCI are based on compressive frame with $B=8$, and those of KH-CVS are based on $B=16$ and two key frames. Please see Visualization 2 for the reconstructed videos.}
\label{fig:exp_main}
\end{figure*}

We used the KH-CVS prototype to record the actual scenes, and the experimental results are shown in Fig. \ref{fig:exp_main}. We compare KH-CVS ($B=16$) with standard SCI ($B=8$) in the same scene. The period of DMD trigger signal is set to 2083 $\mu s$, corresponding to an equivalent frame rate of about 480 frame per second (fps). In KH-CVS, the duration of short exposure $t_x$ is also 2083 $\mu s$ and that of long exposure $t_y$ is set to 33328 $\mu s$. For standard SCI, the exposure time of compressive frames is set to 16664 $\mu s$, which is approximately half to that of KH-CVS, so that the equivalent frame rates in the reconstructed videos of the two systems are kept equal. Note that due to the difference of exposure setups, the brightness of captures are different. In the experiments, the short exposed keyframes are darker than the compressive frames. Before entering the algorithms, they are normalized to keep the same mean value to the intermediate reconstruction frames. This strategy is a common operation that can be also seen in burst high-dynamic-range imaging \cite{hasinoff2016burst}. For reconstruction, GAP-TV, PnP-TV-FastDVDNet, E2E-CNN-8 and RevSCI are used for standard SCI, and the KH-CVS results are reconstructed using the model proposed in this paper (the basic SCI model in KH-CVS is E2E-CNN-16). Based on this setup, first, a rotating fan, with letters marked on several blades of the it, serves as the target scene. The measurements and key frames [Fig. \ref{fig:exp_main} (a)], the reconstructions [Fig. \ref{fig:exp_main} (b)] of standard SCI methods and KH-CVS are shown. In all standard SCI results, the rough outline of the fan blade can be recovered, but the details are relatively blurred and severely distorted by strong artifacts. In the KH-CVS results, the stroke of the character can be clearly recovered. In another scene, we capture a moving toy dinosaur [Fig. \ref{fig:exp_main} (c)-(d)]. Since the object is moved manually and the compressive frames of $B=8$ and $B=16$ are captured one after another, the location of the objects and the motion trajectories are not strictly identical. In the results of standard SCI, the texture of the eye of the toy dinosaur is difficult to resolve with its shape distorted, and the teeth are also blurred. In the results of KH-CVS, the surface texture and edges can be clearly seen with the oval shape of the eye well maintained, and details of the teeth can also be recovered, which also reveals that KH-CVS works well for both bright and dark areas. The results show that the KH-CVS proposed in this paper can achieve better visual quality than standard SCI in real high-speed scenes.

\begin{figure}[!htb]
\begin{center}
\includegraphics[width=0.7\linewidth]{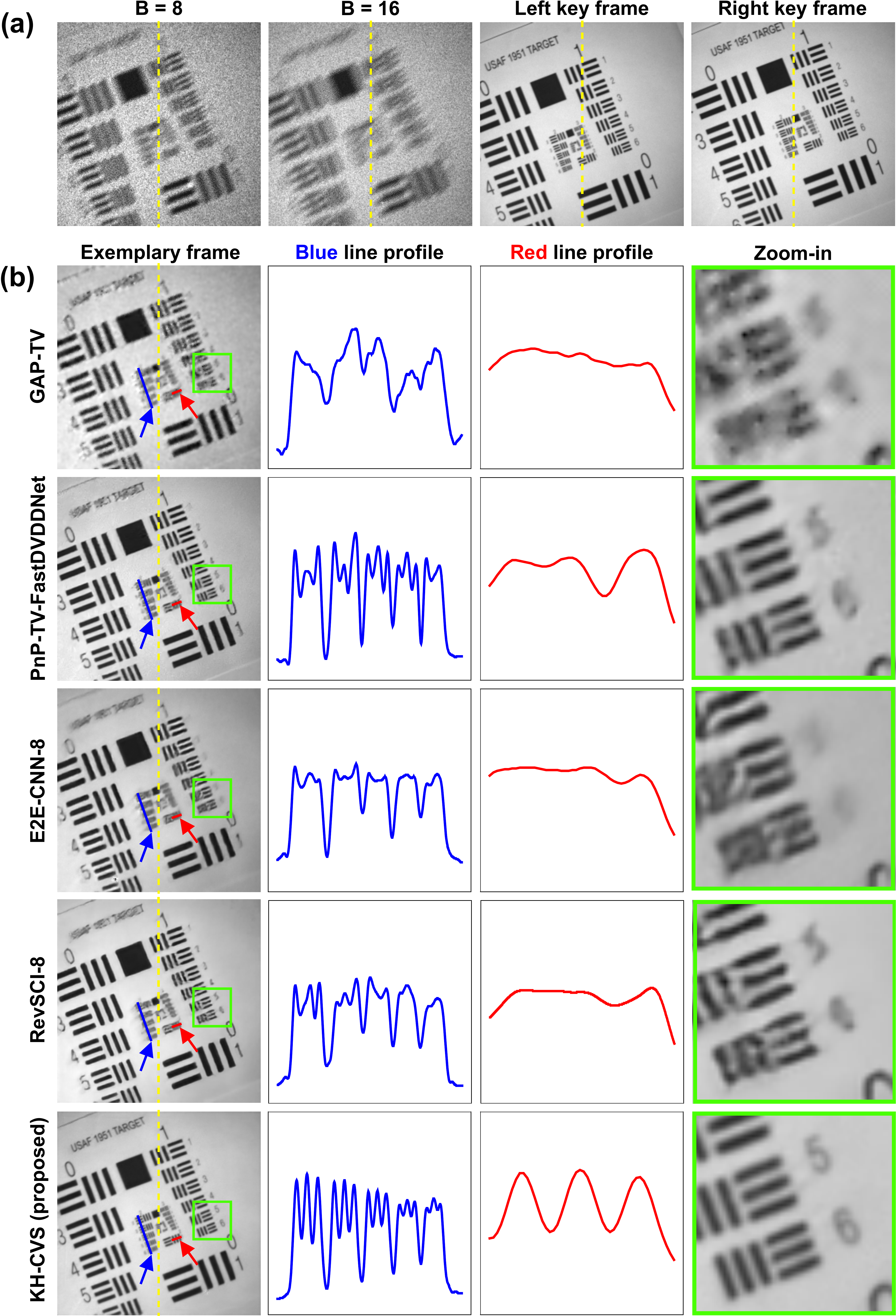}
\end{center}
\caption{Comparison on a moving resolution chart. (a) Compressive frames of $B=8$ and $B=16$, left and right key frames. (b) Exemplary frames reconstructed by different methods. Results of GAP-TV, PnP-TV-FastDVDNet, E2E-CNN-8 and RevSCI are based on compressive frame with $B=8$, and those of KH-CVS are based on $B=16$ and two key frames. Line profiles and zoom-in views are displayed together. The yellow dash lines serve as references to indicate the motion, and the arrows indicate the location of the profiled lines.}
\label{fig:exp_chart}
\end{figure}

To further test the reconstruction quality of KH-CVS, we used a moving USAF 1951 resolution chart as a target. Note that the resolution chart here was printed by us, which is different from its original scale, but can still be used for relative resolution comparison. Measurements for standard SCI and KH-CVS and key frames are shown in Fig. \ref{fig:exp_chart}(a). Exemplary reconstructed frames are shown in Fig. \ref{fig:exp_chart}(b), along with line profiles and zoomed-in views. The blue line plots the horizontal part of group 2, elements 2 to 6, and the red line shows the profiles of group 2, element 1 (vertical parts). For the line profiles, the vertical axis is the grayscale pixel value (from 0 to 255), and the horizontal axis is the normalized spatial distance. The zoomed-in views show the details at group 1, element 5 and 6. In the standard SCI results obtained by the counterpart methods, the bars are affected by strong artifacts and distorted, and the digits are completely blurred. In KH-CVS, they can be clearly distinguished, with better contrast and almost no artifacts.

\section{Discussion and Conclusions}
In summary, KH-CVS offers a new paradigm for compressive video sensing, which brings in benefits to effectively improve the visual quality of reconstructed video. KH-CVS breaks through the standard SCI paradigm that makes reconstruction rely only on a single compressive frame, alternately captures long-exposure compressive frames and short-exposure key frames, and fuses their information to achieve photorealistic compressive video sensing. To this end, we designed a deep neural network architecture that includes basic SCI reconstruction, optical flow estimation, and fusion modules. Through simulation experiments, we verified that KH-CVS has advantages over multiple reconstruction methods \cite{yuan2016generalized,yuan2020plug,zhang2021ten,cheng2020birnat,cheng2021memory,qiao2020deep} of standard SCI with the same level of data amount. Based on our built prototype, we captured actual dynamic scenes at an equivalent frame rate of 480 fps, and the experimental results show that KH-CVS can achieve high visual quality video reconstruction with a wealth of fine details.

KH-CVS is a flexible framework. For the algorithm, we use E2E-CNN as the base model for basic SCI reconstruction, as demonstrated in this paper, but other advanced models can also be integrated into KH-CVS to further improve performance. Also, other optical flow estimators with higher accuracy such as RAFT \cite{teed2020raft} and GMA \cite{jiang2021learning} can be applied. For the model training, we only use a simple L1 loss here, but techniques such as perceptual loss \cite{mathieu2015deep,johnson2016perceptual} and adversarial training \cite{goodfellow2014generative,liu2021single,wang2021generative} can also benefit the perceptual quality. Moreover, for the movements in the scene, we perform warp and fusion based on optical flow here. In addition to flow-based methods, kernal-based methods \cite{choi2019deep,cheng2021multiple} also have shown to be effective in extracting motion spatial information. As SCI has been jointly optimized with vision tasks such as action recognition \cite{okawara2020action}, video object detection \cite{hu2021video} and tracking \cite{hu2021fouriercam,teng2021single}. With the abundant visual information provided by the key frames and the motion information by compressive frames, KH-CVS has the potential to realize more efficient machine vision with lower data bandwidth. For the hardware system, KH-CVS is compatible with deep optics \cite{wetzstein2020inference,zhang2022end} to improve spatial resolution, or it can be adopted to other spatio-temporal compressive imaging systems, such as CUP \cite{gao2014single} and COSUP \cite{liu2021single}. Furthermore, the core idea of KH-CVS is to fuse frames with different exposure properties, thus how to construct more effective sampling strategies is worth further studying. KH-CVS, we believe, will open up new study avenues in the future.

\begin{backmatter}
\bmsection{Funding}
National Natural Science Foundation of China (62135009); The National Key Research and Development Program of China (2019YFB1803500); And by a grant from the Institute for Guo Qiang Tsinghua University.

\bmsection{Disclosures}
The authors declare no conflicts of interest.

\bmsection{Data Availability Statement}
Data underlying the results presented in this paper are not publicly available at this time but may be obtained from the authors upon reasonable request.

\bmsection{Supplemental document}
See Visualization 1 and Visualization 2 for supporting content. 

\end{backmatter}

\bibliography{sample}

\begin{thebibliography}{10}
\newcommand{\enquote}[1]{``#1''}

\bibitem{mait2018computational}
J.~N. Mait, G.~W. Euliss, and R.~A. Athale, \enquote{Computational imaging,}
  {\protect\JournalTitle{Advances in Optics and Photonics}} \textbf{10},
  409--483 (2018).

\bibitem{donoho2006compressed}
D.~L. Donoho, \enquote{Compressed sensing,} {\protect\JournalTitle{IEEE
  Transactions on information theory}} \textbf{52}, 1289--1306 (2006).

\bibitem{yuan2021snapshot}
X.~Yuan, D.~J. Brady, and A.~K. Katsaggelos, \enquote{Snapshot compressive
  imaging: Theory, algorithms, and applications,} {\protect\JournalTitle{IEEE
  Signal Processing Magazine}} \textbf{38}, 65--88 (2021).

\bibitem{llull2013coded}
P.~Llull, X.~Liao, X.~Yuan, J.~Yang, D.~Kittle, L.~Carin, G.~Sapiro, and D.~J.
  Brady, \enquote{Coded aperture compressive temporal imaging,}
  {\protect\JournalTitle{Optics express}} \textbf{21}, 10526--10545 (2013).

\bibitem{koller2015high}
R.~Koller, L.~Schmid, N.~Matsuda, T.~Niederberger, L.~Spinoulas, O.~Cossairt,
  G.~Schuster, and A.~K. Katsaggelos, \enquote{High spatio-temporal resolution
  video with compressed sensing,} {\protect\JournalTitle{Optics express}}
  \textbf{23}, 15992--16007 (2015).

\bibitem{qiao2020snapshot}
M.~Qiao, X.~Liu, and X.~Yuan, \enquote{Snapshot spatial--temporal compressive
  imaging,} {\protect\JournalTitle{Optics letters}} \textbf{45}, 1659--1662
  (2020).

\bibitem{lu2021dual}
R.~Lu, B.~Chen, G.~Liu, Z.~Cheng, M.~Qiao, and X.~Yuan, \enquote{Dual-view
  snapshot compressive imaging via optical flow aided recurrent neural
  network,} {\protect\JournalTitle{International Journal of Computer Vision}}
  \textbf{129}, 3279--3298 (2021).

\bibitem{qiao2020deep}
M.~Qiao, Z.~Meng, J.~Ma, and X.~Yuan, \enquote{Deep learning for video
  compressive sensing,} {\protect\JournalTitle{Apl Photonics}} \textbf{5},
  030801 (2020).

\bibitem{zhang2021ten}
Z.~Zhang, C.~Deng, Y.~Liu, X.~Yuan, J.~Suo, and Q.~Dai, \enquote{Ten-mega-pixel
  snapshot compressive imaging with a hybrid coded aperture,}
  {\protect\JournalTitle{Photonics Research}} \textbf{9}, 2277--2287 (2021).

\bibitem{zhang2022end}
B.~Zhang, X.~Yuan, C.~Deng, Z.~Zhang, J.~Suo, and Q.~Dai, \enquote{End-to-end
  snapshot compressed super-resolution imaging with deep optics,}
  {\protect\JournalTitle{Optica}} \textbf{9}, 451--454 (2022).

\bibitem{wetzstein2020inference}
G.~Wetzstein, A.~Ozcan, S.~Gigan, S.~Fan, D.~Englund, M.~Solja{\v{c}}i{\'c},
  C.~Denz, D.~A. Miller, and D.~Psaltis, \enquote{Inference in artificial
  intelligence with deep optics and photonics,} {\protect\JournalTitle{Nature}}
  \textbf{588}, 39--47 (2020).

\bibitem{yuan2016generalized}
X.~Yuan, \enquote{Generalized alternating projection based total variation
  minimization for compressive sensing,} in \emph{2016 IEEE International
  Conference on Image Processing (ICIP),}  (IEEE, 2016), pp. 2539--2543.

\bibitem{liu2018rank}
Y.~Liu, X.~Yuan, J.~Suo, D.~J. Brady, and Q.~Dai, \enquote{Rank minimization
  for snapshot compressive imaging,} {\protect\JournalTitle{IEEE transactions
  on pattern analysis and machine intelligence}} \textbf{41}, 2990--3006
  (2018).

\bibitem{yang2014video}
J.~Yang, X.~Yuan, X.~Liao, P.~Llull, D.~J. Brady, G.~Sapiro, and L.~Carin,
  \enquote{Video compressive sensing using gaussian mixture models,}
  {\protect\JournalTitle{IEEE Transactions on Image Processing}} \textbf{23},
  4863--4878 (2014).

\bibitem{yang2014compressive}
J.~Yang, X.~Liao, X.~Yuan, P.~Llull, D.~J. Brady, G.~Sapiro, and L.~Carin,
  \enquote{Compressive sensing by learning a gaussian mixture model from
  measurements,} {\protect\JournalTitle{IEEE Transactions on Image Processing}}
  \textbf{24}, 106--119 (2014).

\bibitem{yuan2020plug}
X.~Yuan, Y.~Liu, J.~Suo, and Q.~Dai, \enquote{Plug-and-play algorithms for
  large-scale snapshot compressive imaging,} in \emph{Proceedings of the
  IEEE/CVF Conference on Computer Vision and Pattern Recognition,}  (2020), pp.
  1447--1457.

\bibitem{yuan2021plug}
X.~Yuan, Y.~Liu, J.~Suo, F.~Durand, and Q.~Dai, \enquote{Plug-and-play
  algorithms for video snapshot compressive imaging,}
  {\protect\JournalTitle{IEEE Transactions on Pattern Analysis \& Machine
  Intelligence}} pp. 1--1 (2021).

\bibitem{boyd2011distributed}
S.~Boyd, N.~Parikh, E.~Chu, B.~Peleato, J.~Eckstein \emph{et~al.},
  \enquote{Distributed optimization and statistical learning via the
  alternating direction method of multipliers,}
  {\protect\JournalTitle{Foundations and Trends in Machine learning}}
  \textbf{3}, 1--122 (2011).

\bibitem{liao2014generalized}
X.~Liao, H.~Li, and L.~Carin, \enquote{Generalized alternating projection for
  weighted-2,1 minimization with applications to model-based compressive
  sensing,} {\protect\JournalTitle{SIAM Journal on Imaging Sciences}}
  \textbf{7}, 797--823 (2014).

\bibitem{cheng2020birnat}
Z.~Cheng, R.~Lu, Z.~Wang, H.~Zhang, B.~Chen, Z.~Meng, and X.~Yuan,
  \enquote{Birnat: Bidirectional recurrent neural networks with adversarial
  training for video snapshot compressive imaging,} in \emph{European
  Conference on Computer Vision,}  (Springer, 2020), pp. 258--275.

\bibitem{cheng2021memory}
Z.~Cheng, B.~Chen, G.~Liu, H.~Zhang, R.~Lu, Z.~Wang, and X.~Yuan,
  \enquote{Memory-efficient network for large-scale video compressive sensing,}
  in \emph{Proceedings of the IEEE/CVF Conference on Computer Vision and
  Pattern Recognition,}  (2021), pp. 16246--16255.

\bibitem{ma2019deep}
J.~Ma, X.-Y. Liu, Z.~Shou, and X.~Yuan, \enquote{Deep tensor admm-net for
  snapshot compressive imaging,} in \emph{Proceedings of the IEEE/CVF
  International Conference on Computer Vision,}  (2019), pp. 10223--10232.

\bibitem{jalali2019snapshot}
S.~Jalali and X.~Yuan, \enquote{Snapshot compressed sensing: Performance bounds
  and algorithms,} {\protect\JournalTitle{IEEE Transactions on Information
  Theory}} \textbf{65}, 8005--8024 (2019).

\bibitem{zhang1995video}
H.~Zhang, C.~Y. Low, and S.~W. Smoliar, \enquote{Video parsing and browsing
  using compressed data,} {\protect\JournalTitle{Multimedia tools and
  applications}} \textbf{1}, 89--111 (1995).

\bibitem{feng2020deep}
D.~Feng, C.~Haase-Sch{\"u}tz, L.~Rosenbaum, H.~Hertlein, C.~Glaeser, F.~Timm,
  W.~Wiesbeck, and K.~Dietmayer, \enquote{Deep multi-modal object detection and
  semantic segmentation for autonomous driving: Datasets, methods, and
  challenges,} {\protect\JournalTitle{IEEE Transactions on Intelligent
  Transportation Systems}} \textbf{22}, 1341--1360 (2020).

\bibitem{huang2022deep}
T.~Huang, W.~Dong, J.~Wu, L.~Li, X.~Li, and G.~Shi, \enquote{Deep hyperspectral
  image fusion network with iterative spatio-spectral regularization,}
  {\protect\JournalTitle{IEEE Transactions on Computational Imaging}}
  \textbf{8}, 201--214 (2022).

\bibitem{yuan2015side}
X.~Yuan, T.-H. Tsai, R.~Zhu, P.~Llull, D.~Brady, and L.~Carin,
  \enquote{Compressive hyperspectral imaging with side information,}
  {\protect\JournalTitle{IEEE Journal of Selected Topics in Signal Processing}}
  \textbf{9}, 964--976 (2015).

\bibitem{he2021fast}
W.~He, N.~Yokoya, and X.~Yuan, \enquote{Fast hyperspectral image recovery of
  dual-camera compressive hyperspectral imaging via non-iterative
  subspace-based fusion,} {\protect\JournalTitle{IEEE Transactions on Image
  Processing}} \textbf{30}, 7170--7183 (2021).

\bibitem{pan2019bringing}
L.~Pan, C.~Scheerlinck, X.~Yu, R.~Hartley, M.~Liu, and Y.~Dai,
  \enquote{Bringing a blurry frame alive at high frame-rate with an event
  camera,} in \emph{Proceedings of the IEEE/CVF Conference on Computer Vision
  and Pattern Recognition,}  (2019), pp. 6820--6829.

\bibitem{yuan2018temporal}
M.-Z. Yuan, L.~Gao, H.~Fu, and S.~Xia, \enquote{Temporal upsampling of depth
  maps using a hybrid camera,} {\protect\JournalTitle{IEEE transactions on
  visualization and computer graphics}} \textbf{25}, 1591--1602 (2018).

\bibitem{cheng2021dual}
M.~Cheng, Z.~Ma, M.~S. Asif, Y.~Xu, H.~Liu, W.~Bao, and J.~Sun, \enquote{A dual
  camera system for high spatiotemporal resolution video acquisition,}
  {\protect\JournalTitle{IEEE Transactions on Pattern Analysis and Machine
  Intelligence}} \textbf{43}, 3275--3291 (2021).

\bibitem{paliwal2020deep}
A.~Paliwal and N.~K. Kalantari, \enquote{Deep slow motion video reconstruction
  with hybrid imaging system,} {\protect\JournalTitle{IEEE Transactions on
  Pattern Analysis and Machine Intelligence}} \textbf{42}, 1557--1569 (2020).

\bibitem{reinhard2010high}
E.~Reinhard, W.~Heidrich, P.~Debevec, S.~Pattanaik, G.~Ward, and K.~Myszkowski,
  \emph{High dynamic range imaging: acquisition, display, and image-based
  lighting} (Morgan Kaufmann, 2010).

\bibitem{yuan2017compressive}
X.~Yuan, Y.~Sun, and S.~Pang, \enquote{Compressive video sensing with side
  information,} {\protect\JournalTitle{Applied optics}} \textbf{56}, 2697--2704
  (2017).

\bibitem{sun2017compressive}
Y.~Sun, X.~Yuan, and S.~Pang, \enquote{Compressive high-speed stereo imaging,}
  {\protect\JournalTitle{Optics express}} \textbf{25}, 18182--18190 (2017).

\bibitem{niu2021fast}
B.~Niu, X.~Qu, X.~Guan, and F.~Zhang, \enquote{Fast hdr image generation method
  from a single snapshot image based on frequency division multiplexing
  technology,} {\protect\JournalTitle{Optics Express}} \textbf{29},
  27562--27572 (2021).

\bibitem{ri2006accurate}
S.~Ri, M.~Fujigaki, T.~Matui, and Y.~Morimoto, \enquote{Accurate pixel-to-pixel
  correspondence adjustment in a digital micromirror device camera by using the
  phase-shifting moir{\'e} method,} {\protect\JournalTitle{Applied optics}}
  \textbf{45}, 6940--6946 (2006).

\bibitem{baker2011database}
S.~Baker, D.~Scharstein, J.~Lewis, S.~Roth, M.~J. Black, and R.~Szeliski,
  \enquote{A database and evaluation methodology for optical flow,}
  {\protect\JournalTitle{International journal of computer vision}}
  \textbf{92}, 1--31 (2011).

\bibitem{jiang2018super}
H.~Jiang, D.~Sun, V.~Jampani, M.-H. Yang, E.~Learned-Miller, and J.~Kautz,
  \enquote{Super slomo: High quality estimation of multiple intermediate frames
  for video interpolation,} in \emph{Proceedings of the IEEE conference on
  computer vision and pattern recognition,}  (2018), pp. 9000--9008.

\bibitem{sun2018pwc}
D.~Sun, X.~Yang, M.-Y. Liu, and J.~Kautz, \enquote{Pwc-net: Cnns for optical
  flow using pyramid, warping, and cost volume,} in \emph{Proceedings of the
  IEEE conference on computer vision and pattern recognition,}  (2018), pp.
  8934--8943.

\bibitem{ronneberger2015u}
O.~Ronneberger, P.~Fischer, and T.~Brox, \enquote{U-net: Convolutional networks
  for biomedical image segmentation,} in \emph{International Conference on
  Medical image computing and computer-assisted intervention,}  (Springer,
  2015), pp. 234--241.

\bibitem{niklaus2018context}
S.~Niklaus and F.~Liu, \enquote{Context-aware synthesis for video frame
  interpolation,} in \emph{Proceedings of the IEEE conference on computer
  vision and pattern recognition,}  (2018), pp. 1701--1710.

\bibitem{he2016deep}
K.~He, X.~Zhang, S.~Ren, and J.~Sun, \enquote{Deep residual learning for image
  recognition,} in \emph{Proceedings of the IEEE conference on computer vision
  and pattern recognition,}  (2016), pp. 770--778.

\bibitem{dai2017deformable}
J.~Dai, H.~Qi, Y.~Xiong, Y.~Li, G.~Zhang, H.~Hu, and Y.~Wei,
  \enquote{Deformable convolutional networks,} in \emph{Proceedings of the IEEE
  international conference on computer vision,}  (2017), pp. 764--773.

\bibitem{zhao2016loss}
H.~Zhao, O.~Gallo, I.~Frosio, and J.~Kautz, \enquote{Loss functions for image
  restoration with neural networks,} {\protect\JournalTitle{IEEE Transactions
  on computational imaging}} \textbf{3}, 47--57 (2016).

\bibitem{kingma2014adam}
D.~P. Kingma and J.~Ba, \enquote{Adam: A method for stochastic optimization,}
  in \emph{International Conference on Learning Representations (ICLR),}
  (2015).

\bibitem{russakovsky2015imagenet}
O.~Russakovsky, J.~Deng, H.~Su, J.~Krause, S.~Satheesh, S.~Ma, Z.~Huang,
  A.~Karpathy, A.~Khosla, M.~Bernstein \emph{et~al.}, \enquote{Imagenet large
  scale visual recognition challenge,} {\protect\JournalTitle{International
  journal of computer vision}} \textbf{115}, 211--252 (2015).

\bibitem{kiani2017need}
H.~Kiani~Galoogahi, A.~Fagg, C.~Huang, D.~Ramanan, and S.~Lucey, \enquote{Need
  for speed: A benchmark for higher frame rate object tracking,} in
  \emph{Proceedings of the IEEE International Conference on Computer Vision,}
  (2017), pp. 1125--1134.

\bibitem{wang2004image}
Z.~Wang, A.~C. Bovik, H.~R. Sheikh, and E.~P. Simoncelli, \enquote{Image
  quality assessment: from error visibility to structural similarity,}
  {\protect\JournalTitle{IEEE transactions on image processing}} \textbf{13},
  600--612 (2004).

\bibitem{zhang2018unreasonable}
R.~Zhang, P.~Isola, A.~A. Efros, E.~Shechtman, and O.~Wang, \enquote{The
  unreasonable effectiveness of deep features as a perceptual metric,} in
  \emph{Proceedings of the IEEE conference on computer vision and pattern
  recognition,}  (2018), pp. 586--595.

\bibitem{hasinoff2016burst}
S.~W. Hasinoff, D.~Sharlet, R.~Geiss, A.~Adams, J.~T. Barron, F.~Kainz,
  J.~Chen, and M.~Levoy, \enquote{Burst photography for high dynamic range and
  low-light imaging on mobile cameras,} {\protect\JournalTitle{ACM Transactions
  on Graphics (ToG)}} \textbf{35}, 1--12 (2016).

\bibitem{teed2020raft}
Z.~Teed and J.~Deng, \enquote{Raft: Recurrent all-pairs field transforms for
  optical flow,} in \emph{European conference on computer vision,}  (Springer,
  2020), pp. 402--419.

\bibitem{jiang2021learning}
S.~Jiang, D.~Campbell, Y.~Lu, H.~Li, and R.~Hartley, \enquote{Learning to
  estimate hidden motions with global motion aggregation,} in \emph{Proceedings
  of the IEEE/CVF International Conference on Computer Vision,}  (2021), pp.
  9772--9781.

\bibitem{mathieu2015deep}
M.~Mathieu, C.~Couprie, and Y.~LeCun, \enquote{Deep multi-scale video
  prediction beyond mean square error,} in \emph{International Conference on
  Learning Representations (ICLR),}  (2016).

\bibitem{johnson2016perceptual}
J.~Johnson, A.~Alahi, and L.~Fei-Fei, \enquote{Perceptual losses for real-time
  style transfer and super-resolution,} in \emph{European conference on
  computer vision,}  (Springer, 2016), pp. 694--711.

\bibitem{goodfellow2014generative}
I.~Goodfellow, J.~Pouget-Abadie, M.~Mirza, B.~Xu, D.~Warde-Farley, S.~Ozair,
  A.~Courville, and Y.~Bengio, \enquote{Generative adversarial nets,}
  {\protect\JournalTitle{Advances in neural information processing systems}}
  \textbf{27} (2014).

\bibitem{liu2021single}
X.~Liu, J.~Monteiro, I.~Albuquerque, Y.~Lai, C.~Jiang, S.~Zhang, T.~H. Falk,
  and J.~Liang, \enquote{Single-shot real-time compressed ultrahigh-speed
  imaging enabled by a snapshot-to-video autoencoder,}
  {\protect\JournalTitle{Photonics Research}} \textbf{9}, 2464--2474 (2021).

\bibitem{wang2021generative}
Z.~Wang, Q.~She, and T.~E. Ward, \enquote{Generative adversarial networks in
  computer vision: A survey and taxonomy,} {\protect\JournalTitle{ACM Computing
  Surveys (CSUR)}} \textbf{54}, 1--38 (2021).

\bibitem{choi2019deep}
H.~Choi and I.~V. Baji{\'c}, \enquote{Deep frame prediction for video coding,}
  {\protect\JournalTitle{IEEE Transactions on Circuits and Systems for Video
  Technology}} \textbf{30}, 1843--1855 (2019).

\bibitem{cheng2021multiple}
X.~Cheng and Z.~Chen, \enquote{Multiple video frame interpolation via enhanced
  deformable separable convolution,} {\protect\JournalTitle{IEEE Transactions
  on Pattern Analysis and Machine Intelligence}}  (2021).

\bibitem{okawara2020action}
T.~Okawara, M.~Yoshida, H.~Nagahara, and Y.~Yagi, \enquote{Action recognition
  from a single coded image,} in \emph{2020 IEEE International Conference on
  Computational Photography (ICCP),}  (IEEE, 2020), pp. 1--11.

\bibitem{hu2021video}
C.~Hu, H.~Huang, M.~Chen, S.~Yang, and H.~Chen, \enquote{Video object detection
  from one single image through opto-electronic neural network,}
  {\protect\JournalTitle{APL Photonics}} \textbf{6}, 046104 (2021).

\bibitem{hu2021fouriercam}
C.~Hu, H.~Huang, M.~Chen, S.~Yang, and H.~Chen, \enquote{Fouriercam: a camera
  for video spectrum acquisition in a single shot,}
  {\protect\JournalTitle{Photonics Research}} \textbf{9}, 701--713 (2021).

\bibitem{teng2021single}
J.~Teng, C.~Hu, H.~Huang, M.~Chen, S.~Yang, and H.~Chen, \enquote{Single-shot
  3d tracking based on polarization multiplexed fourier-phase camera,}
  {\protect\JournalTitle{Photonics Research}} \textbf{9}, 1924--1930 (2021).

\bibitem{gao2014single}
L.~Gao, J.~Liang, C.~Li, and L.~V. Wang, \enquote{Single-shot compressed
  ultrafast photography at one hundred billion frames per second,}
  {\protect\JournalTitle{Nature}} \textbf{516}, 74--77 (2014).

\end{thebibliography}
\end{document}